# Structural insights into characterizing binding sites in EGFR kinase mutants

Zheng Zhao[1], Lei Xie[2,3] and Philip E. Bourne*[,1,4]

*1*. Department of Biomedical Engineering, University of Virginia, Charlottesville, Virginia 22904, United States of America

*2*. Department of Computer Science, Hunter College, The City University of New York, New York, New York 10065, United States of America

*3*.The Graduate Center, The City University of New York, New York, New York 10016, United States of America

*4*. Data Science Institute, University of Virginia, Charlottesville, Virginia 22904, United States of America

**\*Corresponding author**

   Philip E. Bourne: phone, (434) 924-6867; e-mail, peb6a@virginia.edu

# Abstract

Over the last two decades epidermal growth factor receptor (EGFR) kinase has become an important target to treat non-small cell lung cancer (NSCLC). Currently, three generations of EGFR kinase-targeted small molecule drugs have been FDA approved. They nominally produce a response at the start of treatment and lead to a substantial survival benefit for patients. However, long-term treatment results in acquired drug resistance and further vulnerability to NSCLC. Therefore, novel EGFR kinase inhibitors that specially overcome acquired mutations are urgently needed. To this end, we carried out a comprehensive study of different EGFR kinase mutants using a structural systems pharmacology strategy. Our analysis shows that both wild-type and mutated structures exhibit multiple conformational states that have not been observed in solved crystal structures. We show that this conformational flexibility accommodates diverse types of ligands with multiple types of binding modes. These results provide insights for designing a new-generation of EGFR kinase inhibitor that combats acquired drug-resistant mutations through a multi-conformation-based drug design strategy.

# Introduction

Epidermal growth factor receptor (EGFR) is one of four members of the transmembrane epidermal growth factor receptor family (EGF/ErbB receptor family). It is typically activated by an epidermal growth factor (EGF) that binds to the extracellular domain which protrudes from the cell membrane thereby regulating signaling pathways[1]. Mutation of the EGFR kinase domain leads to higher activity thereby stimulating four major downstream signaling pathways including MAPK/ERK, PLCγ/PKC, JAK/STAT and PI3K/AKT which impact transcription and cell cycle progression[2] leading to cancers and inflammatory diseases[3]. Specifically, it has been shown that diverse mutations in the EGFR kinase domain, typically L858R and del746-750, are associated with non-small cell lung cancer (NSCLC) which accounts for about 80% to 85% of all lung cancer cases[4]. Thus, EGFR kinase inhibitors that target these activation mutations are desirable for the treatment of NSCLC. Since 2009, five EGFR kinase inhibitors have been approved by the US Food and Drug Administration (FDA). The first two, Gefitinib and Erlotinib, show efficacy at the start of treatment. However, after about 12 months, drug resistance from an acquired T790M mutation arises[5]. Thus, a second-generation EGFR kinase inhibitor, Afatinib, was developed for the T790M mutation. Because of the limited therapeutic potential of Afatinib[6], soon after, a third-generation of EGFR kinase inhibitors, Osimertinib and Olmutinib, were developed. Afatinib, Osimertinib and Olmutinib are all irreversible inhibitors[7] that form a covalent bond with Cys797. However, the occurrence of an acquired C797S mutation greatly reduces the efficacy of these three covalent drugs. Therefore, to date, this acquired resistance remains a major challenge in the treatment of NSCLC[8-9]. New EGFR kinase inhibitors that overcome these acquired mutations are needed. Recently, Jia et al. discovered an allosteric EGFR kinase inhibitor (EAI045)[10], which overcomes both T790M and C797S mutations, and offers a means for treating NSCLC. Further, a few non-covalent ATP-competitive inhibitors that bind to the ATP binding site were rationally designed to overcome acquired resistance[11-15]. For example, Marcel Günther et al.[14] developed a reversible inhibitor that can inhibit the L858R/T790M/C797S triple mutant. Hwangseo Park et al.[15] also identified a few ATP-competitive inhibitors that overcome the Del746-750/T790M/C797S/ triple mutant. This next-generation allosteric or non-covalent inhibitors demonstrate promising potential and an opportunity to overcome the multiple mutations through designing elaborate inhibitors that match the mutated binding site of the EGFR kinase[16]. These findings prompted us to revisit the EGFR kinase binding site to obtain a detailed understanding of the similarity and differences between the wild-type and mutant EGFR kinases.

Currently, with many available EGFR structures, there is the data-driven opportunity to focus on the EGFR binding site using a structural systems pharmacology strategy[17-20]. We collected all released structures of the EGFR kinase domain to analyze the overall conformational space of the binding sites as well as the respective kinase-ligand binding modes. We focused on all diverse acquired mutations at the binding sites, exploring the binding site flexibility of EGFR kinase domains with the del746-750/T790M/C797S/ mutation and the L858R/T790M/C797S mutations by using µs-scale molecular dynamics (MD). MD simulation has been successfully applied to study the EGFR conformation space at a computational physiological environment[21-24]. For example, Park et. al explored the EGFR conformational transition between the inactive and active state to determine the role of gatekeeper mutation on inhibitor selectivity using MD simulations[21]. Kannan et. al used the combined MD simulation with binding assay data to reveal a mutant specific pocket[24]. Additionally, we scrutinized the binding site features at an atomic level using the function-site interaction fingerprint approach and the volumetric analysis of surface properties. Our analyses show both wild-type and mutated structures encompass multiple conformational states. Novel conformational state that have not been observed in the solved structures can exist. The conformational flexibility in the structure accommodates diverse types of ligands with multiple types of binding modes. These results provide us with critical insights into future drug design in treating drug resistant NSCLC and other cancers[25].

# Results

**1. EGFR kinase conformations and mutations**

The similarities and differences in EGFR kinase binding sites can be determined by comparing their X-ray structures. We clustered EGFR kinase structures in our dataset into two kinase states, active and inactive and into six classes (Figure 1a). In these classes, the side chain of the aspartic acid in the DFG motif points in three different directions: "DFG-In", "DFG-Out", and DFG-$\frac{1}{2}$In. In DFG-$\frac{1}{2}$In, the side chain of the aspartic acid points to the roof[26]. Similarly, the C-helix also exhibits significant displacements called "In" and "Out", and the state between them (labeled "$\frac{1}{2}$Out"). The number of x-ray structures and the representative structure in each class are shown in Figure 1a (a complete clustering of structures is in Table S1). Using the representative structure from each class, the major differences in the P-loop, C-helix and DFG motifs are highlighted in Figure 1b. In the active state (Class-1 and Class-2), the DFG motif is in a "DFG-In" state. The C-helix is found in the "In" state (class-1, Figure 1a-b) or has a small displacement referred to as "$\frac{1}{2}$Out" (Class-2, Figure 1a-b). In the inactive kinase (class-3-6), the DFG motif is found in three directions and the C-helix is found in the "Out" state in all structures. These distinctly different conformations in the binding site of EGFR kinases facilitates the identification of potential sites outside of the conserved ATP-binding site, providing a structural basis to accelerate structure-based drug design through the discovery of novel binding modes.

Acquired multi-drug resistance motivates the search for drugs that can target mutated structures. To begin this process, we catalog the currently available conformations of EGFR kinase domains, both wildtype and mutations, according to the scheme described above (Figure 1c). The wildtype (WT) EGFR kinases (37 structures) fall into four classes of conformation (Class-1,3,4,5). The L858R/T790 mutation (41structures) also exhibits different conformational states (Class-1,2,3,5,6). Other mutated classes such as T790M, which have a limited number of released structures (14 structures), present fewer unique conformations (Class-1,2,3). As catalogued, the WT and mutated structures can have the same conformation. For example, three types of EGFR kinase, the WT, L858R and T790M/L858R (Figure 1c) all adopt the Class-5 conformation. In summary, WT and mutated structures both present multiple conformations. From a drug design perspective, the good news is that multiple conformations suggest flexibility in the EFGR kinase domain providing opportunities to design potent and specific inhibitors by accommodating a unique conformation. The bad news is the challenge to protein-ligand docking and modeling quantitative structure-activity relationship since the ligand-induced conformation is not known a priori. Here we attempt to address this shortcoming.

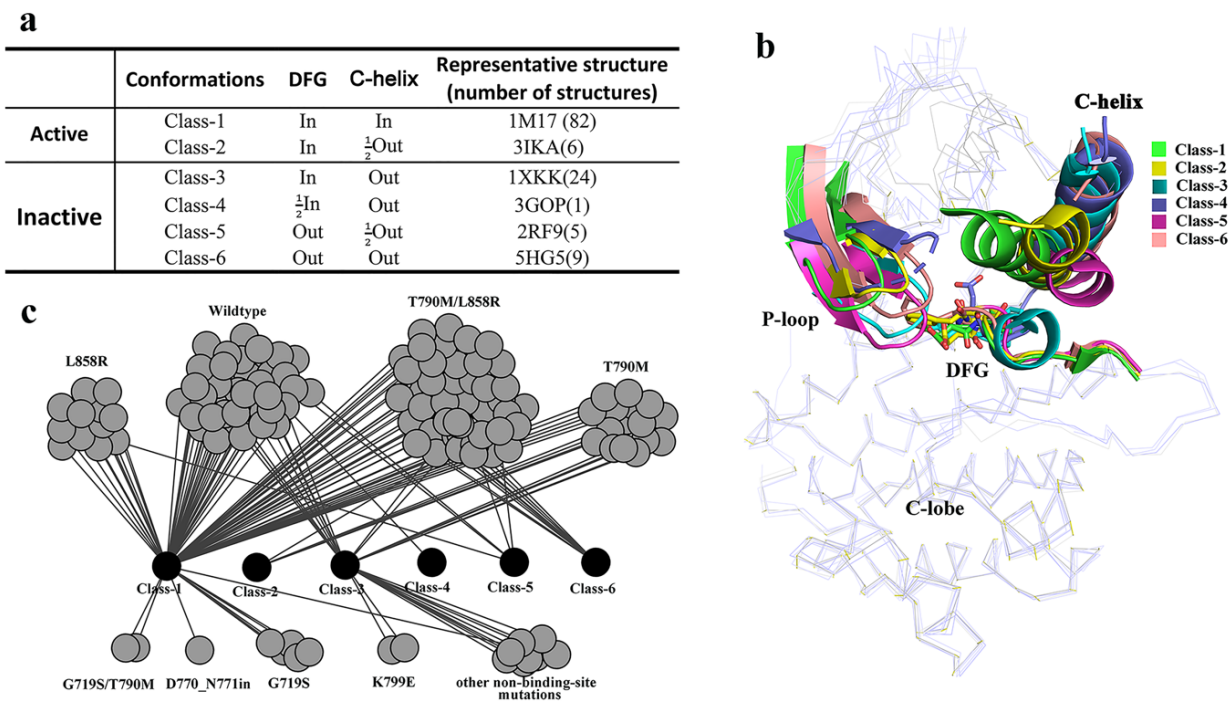

**Figure 1.** EGFR kinase structure dataset with mutations. (a) Classes of EGFR kinase domain conformations. (b) Comparison of different classes of EGFR kinase with representative conformations. (c) Network of currently released structure conformations and mutations.

We also marked two important motifs, the R- and C-spine[27-29] (Figure S1). The catalytic C-spine is assumed highly stable since it is strikingly similar in all six classes of EGFR states. However, the R-spine has different architectures. In Class-1 and 2 at the active state, the R-spine residues are linearly connected. In Class-3 and 4 at the DFG-in/C-helix-out inactive state, the R-spine is partially assembled because the C-helix displacement leads to M766 movement. In the DFG-out inactive state (Class-5 and 6), the R-spine is not assembled because the sidechain of the F856 residue flips to the other side of the DFG peptide.

## 2. Comparing binding sites among EGFR wildtype and mutants

We calculate changes in the size and shape of the binding pockets for different classes of conformation. The volume of the binding pockets is 1119 Å$^3$, 953 Å$^3$, 950 Å$^3$, 1056 Å$^3$ and 972 Å$^3$ for Class-1 to 5, respectively (Figure S2). These pockets are not significantly different, however, the volume of the binding pocket in Class-6 is significantly increased (1913 Å$^3$). It can thus accommodate more diverse types of small molecule ligands. The difference in the shape and size of the binding pockets is shown in Figure 2. Given that the Class-1 conformation (Figure 2a) is the most frequently observed (Figure 1a), we compared the binding cavities of the other five classes with that of Class-1. Specifically, we decomposed the binding cavity into five sub-pockets: front pocket (FP), GxGxxG motif at the P-loop (G-rich-loop), and back clefts BP-I, BP-II, and BP-III[30], and quantitatively compared them (Figure 2b-f (upper and lower). BP-I is located at the back cleft close to the adenine pocket; BP-II includes the hydrophobic pocket at the back cleft; BP-III is the sub-pocket at the back cleft comprising the DFG peptide, C-terminal end of the C-helix and the N-terminus of the catalytic loop. Table S2 lists the volume of every sub-pocket.

Firstly, compared with the back cleft of Class-1, the volume of the back cleft in Class-2 and Class-5 is smaller. Their back binding sub-pocket is not formed (Figure 2b and 2e). However, the BP-II cleft of Class-3, the BP-II cleft of Class-4 and the BP-II, III clefts of Class-6 is present (Figure 2c,2d and 2f). It is worth noting that the C-helix in Class 2-6 is in "OUT" or "$\frac{1}{2}$Out" state. Thus, the C-helix displacement does not always induce the formation of the back cleft. Secondly, in Class 1-6, the volumetric size of the FP region that binds ATP is similar but the shape is slightly different (Figure 2 and Table S2). Finally, for the sub-pocket at the G-rich loop, a large binding pocket is formed in Class-2, Class-5 and Class-6. However, there is no space to accommodate a ligand in Class-3 and Class-4 similar to Class-1 (Figure 2b-f). Interestingly, Class-6 (Figure 2f) is different from other classes; besides the regions of FP and G-rich-loop, a large sub-pocket is formed at BP-II and BP-III. Thus, the cavity of Class-6 has adequate space, not only at the ATP-competitive site, but also at the allosteric site. This conformational mode may provide the structural basis for the design of diverse type-II and/or type-III EGFR kinase inhibitors[31-32].

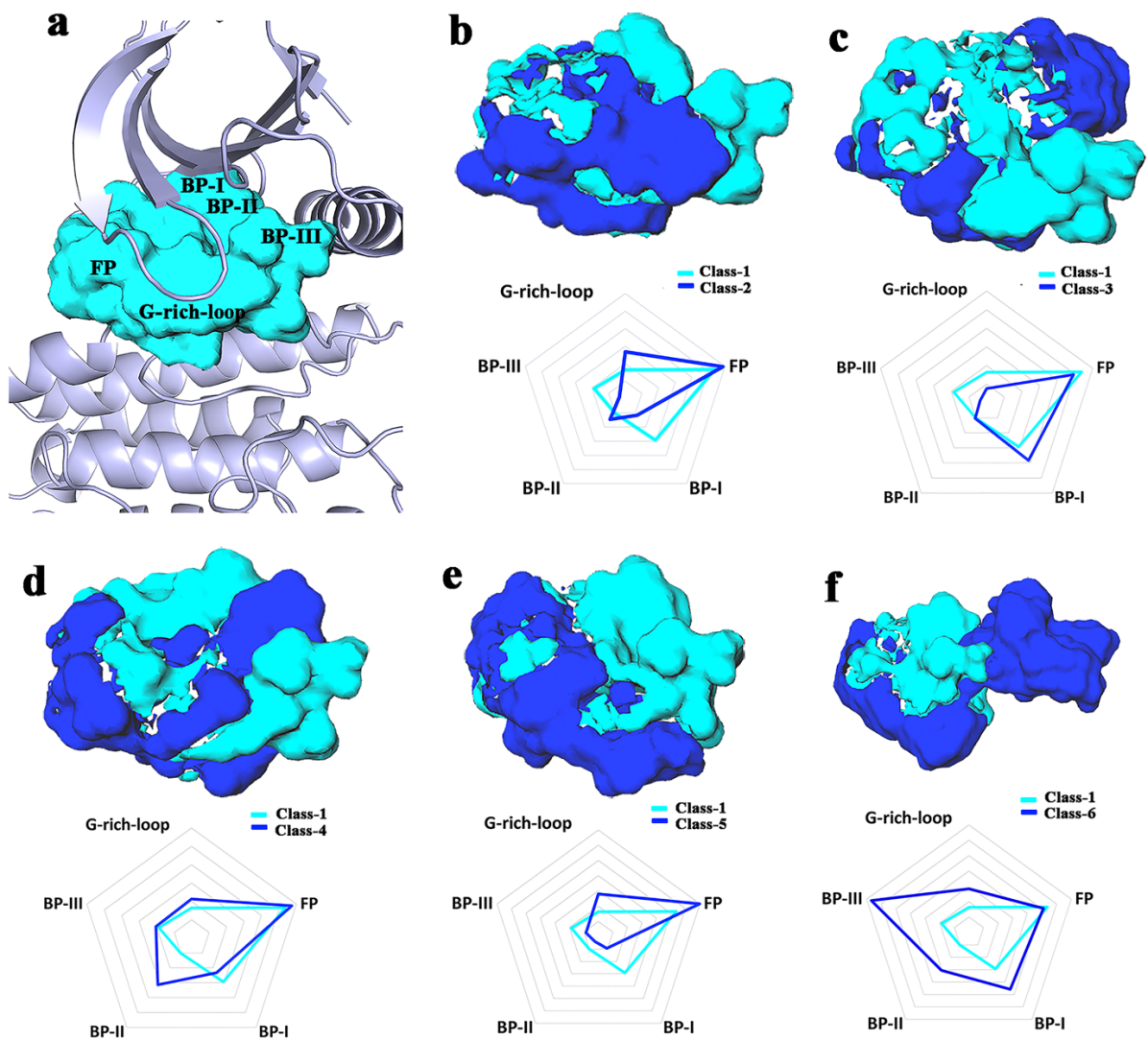

**Figure 2.** Differences in binding pockets. The cavity of Class-1 is showed in cyan and the cavity of each other class in blue, respectively. (a) The Class-1 binding pocket shown in PDB 1m17 as a reference showing every region. (b-f) Significant pairwise differences (upper); quantitative

difference of every pairwise binding pocket illustrated using a radar chart and a contour interval of 100 Å$^3$ (lower).

In summary, we compared the different binding pockets quantitatively and qualitatively. The changes in the conformations in the C-helix and DFG motif induces diverse sub-pockets within the binding cavities. It is worth noting that we only analyze the difference in the binding cavities using the rigid x-ray structures. Nevertheless, the induced sub-pockets present the diverse shape and size of the binding pockets. It suggests that the plasticity of the EGFR kinase binding pocket combined with the inducing dynamic conformational space introduced by molecular dynamics has great potential for designing inhibitors with desired selectivity profiles.

### 3. Dynamic conformational space of the binding sites

Starting from their X-ray structures, the μs-scale equilibration of the trajectories yields a large number of conformations for the wildtype and different mutated EGFR kinases. The trajectory analysis provides information on the similarity of conformations and frequent conformational transitions based on the total energy curves (Figure S3), the root mean square deviation of Cα atoms (Cα-RMSD) (Figure S4) and the root mean square fluctuation of Cα atoms (Cα-RMSF) (Figure S5). The main collective motions, as inferred from the first principal component of PCA, are shown in Figure S6. For all three systems, their ATP binding pockets show little variation. Around the binding site, the flexibility mainly comes from P-loop, the activation loop and C-helix. From specific detailed comparison of the binding sites, these conformations show a high degree of deviation from the initial crystal x-ray structure (PDB id: 1m17, Class-1, Figure 1a). Starting from the DFG-In/C-helix-In conformations, the resulting equilibrated structures present different conformations including both the active and inactive states (Figure 3). Among them, the DFG motif exhibits significant displacement from DFG-In to DFG-out, and the C-helix transitions from C-helix-in to C-helix-out. In the trajectory of wild-type structures (Figure 3a), all six classes of conformation can be sampled. As shown in Figure 3b-c for the mutated structures of del746-750/T790M/C797S and L858R/T790M/C797S, the conformations range from active state (Class-1 and Class-2) to inactive (Class-3 and Class-4). It is well known that conformational transition needs to overcome the free energy barrier[33]. It could be the reason that we cannot capture Class-5 and Class-6 conformations using unbiased MD simulation given our current MD time-scale. To explore a larger conformational space improved sampling technologies or longer simulation time are required[22-23] as described by Shan[22], Sutto[23], and Park[21]. Nevertheless, it is clear, even from traditional MD simulation, that multiple conformational states of the EGFR kinase in both WT and mutated systems exist. Importantly, we observe a new conformational state (labeled as Class-D1 in Figure 3c), which doesn't occur in the EGFR kinase structure dataset. This new conformation (Figure 4) has "DFG-$\frac{1}{2}$In/C-helix-out", in which the DFG flipping and C-helix-out displacement are similar to that of Class-4. The difference is that the salt bridge between K745 and E762 occurred in Class-D1 and a hydrogen-bond interaction formed between D855 and K745. This salt bridge and the hydrogen-bond interaction stabilize this dynamic conformation (Figure 4). In the Class-D1 state, when C-helix displacement occurs, the sheets consisting of beta-strands 1-5 have a combined displacement outward (Figure 4). It will be interesting to screen for novel inhibitors using this new conformation.

We further analyzed the features of the Class-D1 binding pocket. The volume of the binding pocket is 1204 Å$^3$ (Figure S7), which is slightly larger than that of Class 1-5. We compared the Class-D1 binding site with the other six classes of binding sites. The main difference between Class-D1 and Class 1-5 is the BP-III region. The main difference from Class 6 is the G-rich-loop region (Figure S8). Obviously, the Class-D1 binding pocket has a large sub-pocket at the BP-III region. More specifically, we calculated the similarity between the Class-D1 binding pocket and the other six classes of binding pockets using the volumetric distance[34] (Table S3), which shows the Class-6 binding site is most similar to Class-D1. Like Class 6, the slightly larger BP-III sub-pocket is also available in the Class-D1 state. Additionally, to estimate the possible binding mode, we screened the whole kinase structurome, including 3180 kinase-ligand structures[35], to obtain the most similar binding pocket to the Class-D1 binding pocket. The top complex (Table S4) is a JAK1 kinase structure[36] (pdb id: 4e4n), where the ligand (0NL) binds to the pocket by bearing a heterocycloamine scaffold, forming hydrogen bonds with the hinge peptide and a t-butyl carbamate at the G-rich-loop region (Figure S9). We also noted that the distance between the t-butyl and the C-helix is 9.6 Å, which means this BP-III sub-pocket region is not fully utilized. This binding mode for the JAK1-0NL complex can guide new EGFR inhibitor design through binding the Class-D1 conformation. The complete structure file (PDB format) of this new conformation is available as Supporting Information.

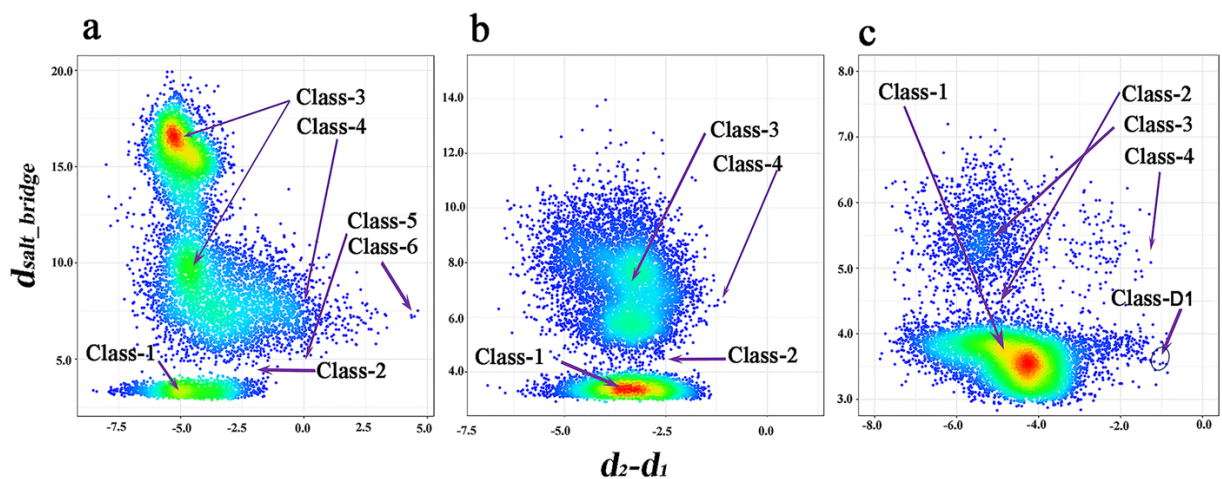

**Figure 3.** Scatter density plot of the distributions of conformation space for the different systems (a) wildtype, (b) del746-750/T790M/C797S and (c) L858R/T790M/C797S. The y axis represents the distance of K745 Nζ : E762 Cδ and the x axis represents the flexibility of the DFG peptide. $d_2$-

$d_1=d_2$[C797 Cα : D855 Cγ]–$d_1$[C797 Cα : F856 Cζ]. High-density and low-density regions are colored in red and blue, respectively.

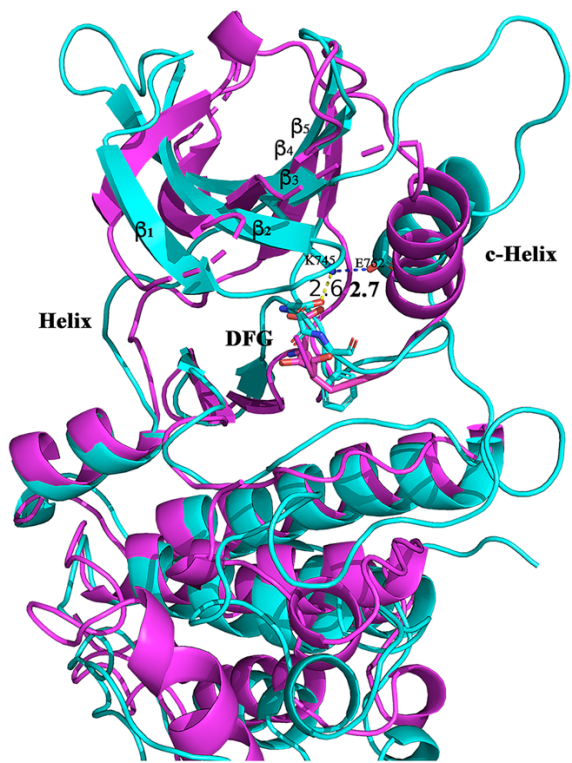

**Figure 4**. The Class-D1 of conformation (cyan color) comparing with that of Class-4 in purple color.

In summary, the MD simulation reveals more conformational states than those in the released x-ray structure dataset. It is important to analyze the multiple conformational states in the context of the diverse inhibitors that bind to those different conformations. Stated another way, multiple conformations will facilitate novel inhibitor design with a specific SAR. Note, however, that the ligand could also induce a new EGFR kinase conformation with a yet to be discovered binding pocket[37] which points to the need for further structure determination of new ligand-bound complex structures, as well as further computational studies to sample a larger conformation space in the process of drug design in order to obtain the desired SAR[38].

## 4. EGFR kinase-ligand binding characterization

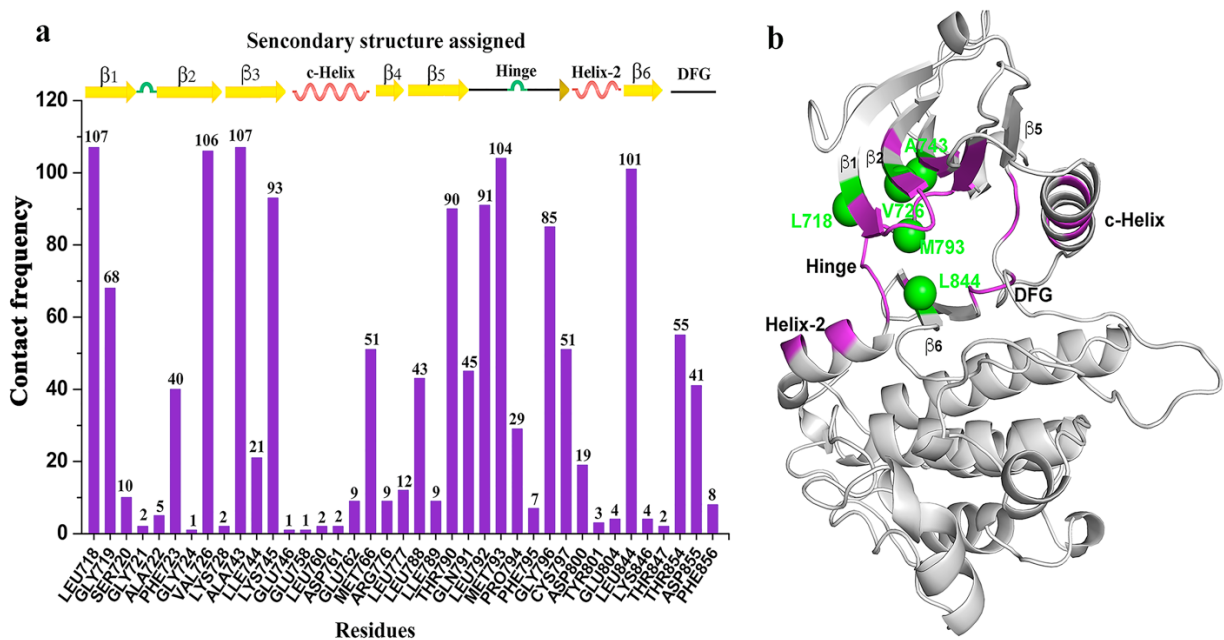

**Figure 5.** Interaction details between the amino acids that constitute the binding site and the corresponding ligand. (a) the contact frequency of each amino acid providing the binding interaction and the corresponding locations. (b) distribution of interacting amino acids in the 3D binding site (purple) and the top 5 residues with the highest contact times (green).

It is well known that the kinase ATP binding site is highly conserved across the whole kinome. In order to achieve the desired selectivity, it is critical to understand the atomic details of kinase-inhibitor interactions[35]. Here, using all released EGFR kinase crystal structures, we analyze and characterize the atomic details of interactions between the ligand and the residues that constitute the EGFR kinase ATP binding site (see Methods and Table S5 for specific PDB structures). Figure 5 shows the contact frequency between specific residues and the bound ligand. Seven types of interatomic interaction are defined based on geometric rules [39] but not energy (see Methods). The analysis identifies 39 amino acids that are close to the binding site (Figure 5a). The 39 residues are located predominantly in beta-sheets or helixes as shown in Figure 5a and 5b. The residues with the top 5 contact frequencies are L718, V726, A743, M793 and L844. They are located at β1-3, Hinge and β6 (Figure 5b in green) and form a core binding pocket. The core binding pocket is highly hydrophobic, largely conserved and consistent with previous results[35]. The core binding pocket accommodates the adenine base of ATP or the ATP-competitive inhibitor. The high contact frequencies at the core binding pocket also suggests that most of the co-crystallized EGFR kinase inhibitors are ATP-competitive. However, by selectively binding other residues with lower contact frequency, higher selectivity can be anticipated.

To reveal the nature of binding specificity, we compare the pairwise similarity of aligned residue-ligand functional site interaction fingerprints (Fs-IFPs) (see Methods). Hierarchical clustering of the Fs-IFPs for all EGFR kinase-ligand co-crystallized structures was conducted and the binding modes grouped into six clusters that represent different binding modes (Figure 6a and Table S6). In cluster-1 (Figure 6b), the ligand is located at the position of the ATP binding site, where the core binding pocket is highlighted as aforementioned. It is not surprising that most inhibitors fall into this cluster. Typically, there are two or three hydrogen-bond interactions between the ligand and the amino acids located at the hinge[31]. Since the core binding pocket is conserved, this type of inhibitor has a multi-targeted effect across the kinome. For example, the ligand STO (1,2,3,4-tetrahydrogen staurosporine) (Figure 6b) is a derivative of the natural product staurosporine[40], which is known to have broad cross-reactivity[41]. Following the cross reactivity observed, the ligand STO has a specific kinase spectrum that can be targeted. By combining the target spectrum of a given compound, further SAR optimization becomes a valid strategy[26, 42]. By using this strategy, a close analogue of staurosporine (PKC-412, a multi-kinase inhibitor) was approved in April 2017 for treating acute myeloid leukemia (AML), myelodysplastic syndrome (MDS) and advanced systemic mastocytosis[43-45]. Similar to cluster-1, the cluster-2 interaction patterns also include the ATP-competitive binding mode (Figure 6c). However, in this cluster the kinase is in the inactive state and belongs to the Class-6 conformation as described in Figure 1. In this case the ligand does not have any interaction with the C-helix. Rather the benzenesulfonyl fluoride ligand fragment presents characteristic interactions, namely the aromatic π-π stacking interaction with the phenylalanine side chain of the DFG motif and the electrostatic interactions with the amino acids on the G-rich loop. In cluster-3 (Figure 6d), the ligand also presents an ATP-competitive binding mode. However, different from cluster-1, the ligand has a hydrophobic fragment penetrating deeply into the hydrophobic pocket that adjoins the core binding pocket and is located at the back of the kinase binding site[30]. The hydrophobic pocket is often used to achieve the desired selectivity, but mutation of the gatekeeper residues directly affects the efficacy of this type of inhibitor[46]. Cluster-4 (Figure 6e) has a non-ATP-competitive binding mode. The ligand is located at the back of the binding site, and occupies the hydrophobic pocket, similar to cluster-3, as well as the allosteric site[47]. In this cluster, the C-helix is displaced outward to the "Out" state and opens up the allosteric site. Thus, the ligand forms interactions with the DFG motif and the C-helix. The allosteric binding mode provides high selectivity overcoming the T790M/C797S mutations[10]. In cluster-5 (Figure 6f), ligand binding presents two features. Firstly, the ligand has a phenyl group to bind the hydrophobic pocket that is similar to cluster-3 and cluster-4. Secondly, the tail of the ligand interacts with not only the sidechain of aspartic acid from the DFG motif but also the sidechain of asparagine 842 at $\beta_6$. It is notable that the position of the tail is similar to that in cluster-2, but the binding mode is totally different. Firstly, the tail of the ligand forms an interaction with the aspartic acid of DFG-in in cluster-5, but with the phenylalanine of DFG-out in cluster-2. Secondly, the tail extends towards the A-loop and forms an interaction with Asn842 in cluster-5, but extends towards the P-loop and is in contact with the G-rich loop in cluster-2. In cluster-6, the ligand is an ATP-competitive binder. The furan group of the ligand not only forms very subtle interactions with the hydrophobic pocket but also forms polar interactions with K745, E762 and the DGF motif. As aforementioned, in cluster-4, the C-helix contributes to the binding affinity of the allosteric inhibitor. However, although the inhibitors in cluster-6 are not allosteric inhibitors, the C-helix is in the "C-helix-In" state in close proximity to the oxygen atoms of the furan group thereby forming polar interactions. Thus, the furan group provides unique binding characteristics.

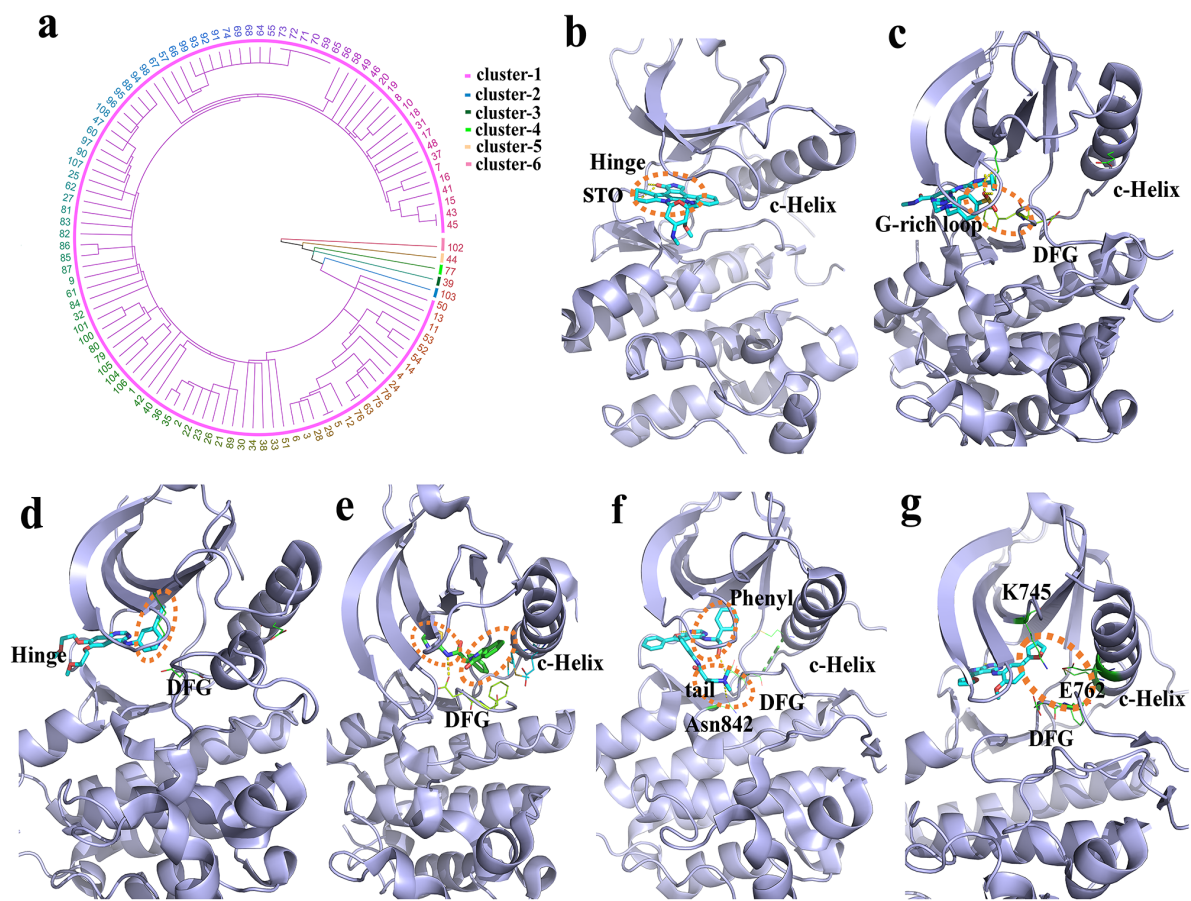

**Figure 6**. EGRF ligand-binding clusters and features. (a) hierarchical cluster analysis. The number represents the index of every structure (Table S5). (b-g) the clustered binding modes (clusters 1-6) are shown using the PDB structures 2itu, 5u8l, 4hjo, 5d41, 4jq8 and 4jeb, respectively. The dash-line circles (orange) highlight the binding characteristics of the core binding pocket in: (b) the location of binding G-rich loop and DFG motif; (c) the hydrophobic pocket; (d) the hydrophobic pocket and the allosteric site; (e) the hydrophobic pocket and the location of binding DFG and Asn842; (f) and the combined binding position with the interactions from the hydrophobic pocket and the allosteric site (g).

In summary, the binding modes of co-crystallized EGFR kinase-ligand complexes are diverse. They provide the structural basis for mediating inhibitor selectivity, important to keep in mind when designing mutant-sensitive inhibitors.

## Conclusions

We have explored the impact of mutations on structural conformation of the EGFR kinase domain for all released crystal x-ray structures. From a detailed structural analysis, we identified six classes of conformation which focus on the flipping of the DFG motif and the displacement of the C-helix. There is no apparent correlation between these mutations and conformation. For example, the common T790M mutation adopted not only the typical DFG-in/C-helix-in conformation but also the DFG-out/C-helix-out. We suggest that the conformations of the EGFR kinase are flexible and can fall into distinct locally stable states. From the perspective of drug design, multi-conformation-based drug screening should be of significant value when targeting EGFR kinases.

We also explored the dynamic conformation space occupied by the EGFR kinase using μs-level MD simulation. We not only sampled the multiple different conformations from Class-1 to Class-6 but also found a new conformation mode. In the two acquired mutated kinase systems, we did not sample Class-5 and Class-6 conformations; advanced sampling technologies are needed. Less of an issue perhaps since Class-5 and Class-6 conformations that were obtained from the released ligand-bound structures (Figure 1b), as well as taking conformational flexibility into account, suggests that the EGFR kinase-ligand binding would induce a specific binding mode and EGFR kinase conformation change. Recently, Sonti et al.[38] reported a similar ligand-induced binding mechanism for Abelson tyrosine kinase (Abl). Abl belongs to the TK group, like the EGFR kinase, and has 60% sequence similarity[48]. Therefore, it is important for EGFR kinase-ligand SAR to take the induced binding mechanism into consideration. This presents a challenge to *in silico* compound screening for it requires the protein target be considered not as a static structure but as an ensemble of conformations[49].

Such a change in conformation often results in a change in the binding site. We found that the outward displacement of the C-helix does not always form the BP sub-pockets, (Figure 2b and 2e). Moreover, with these conformational changes the volume and shape of the binding cavities shows systematic plasticity (Figure 2b-f), which can be further elucidated into distinctive binding patterns through binding diverse inhibitors. The derived Fs-IFPs highlight this distinctive protein-ligand binding patterns at the atomic level. It is our hope that within each sub-cluster of fingerprints, the unique binding modes will provide useful clues in designing unique molecular fragments for a new series of EGFR kinase inhibitors.

# Method

## 1. Dataset of EGFR kinase structures

We collected all 127 PDB structures of the EGFR kinase domain released through Dec. 2017 from the Protein Data Bank ( www.rcsb.org )[50] by using the human EGFR UniProt[51] identifier, P00533. There are 20 apo and 107 holo PDB structures in the dataset. The binding sites of all structures were aligned using the SMAP software[52-54]. We clustered all structures based on the Cα-RMSD of the EGFR N-terminal lobes. The mutations closest to the binding site of each structure was extracted to determine the relationships between the conformation patterns and the mutations.

## 2. Differentiating the binding sites

To differentiate the different binding sites, first the given binding sites were pairwise aligned using the SMAP tool[54]. Then, each binding site and each sub-pocket was detected using the Surflex software[55-56] with default proto parameters. For each sub-pocket, a set of residues bordering the sub-pocket were chose based on Liao's description[30]. Then the volumetric size and shape of each binding site or each sub-pocket was inferred using the volumetric analysis of surface properties (VASP) software[34], in which all computations were carried out at 0.5 A resolution. Lastly, differentiating the binding sites was determined using the volumetric difference of constructive solid geometry[34]. The similarity was calculated by using the volumetric distance $v_{(x,y)}$,

$$v_{(x,y)} = 1 - \frac{v(x \cap y)}{\min(v(x), v(y))}$$

where, $v(i)$ represents the volume of a given binding site $i$ in Å$^3$. $v(x \cap y)$ represents the volumetric intersection of $v(x)$ and $v(y)$ for x and y are binding pockets.

## 3. Sampling the conformational space using MD

We carried out microsecond-level molecular dynamics (MD) simulation for three systems, the wild-type and two acquired mutated types (L858R/T790M/C797S and del746-750/T790M/C719S). First, the initial structure was prepared from PDB 1m17 with the DFG-in/C-helix-in active conformation as the wild-type EGFR kinase domain. Then the corresponding mutated models were built for each system, respectively, using PDB 1m17 as the template and the software modeller[57]. Then each structure was solvated in a cubic box of water molecules with 12Å distance from the solvated molecules and neutralized by adding $Cl^-$ and $Na^+$ using an ACEMD setup script[58]. Finally, a general short-MD protocol was used for initial minimization and equilibration including 2ps minimization, 100ps for NVT, 1ns for NPT with heavy-atom constraints and 1ns for NPT without any constraints. The next stage was a 1.5 μs equilibration process without the constraint and the last 1000ns of MD trajectories were retained for further analysis.

The MD simulations were carried out using the ACEMD software[58] with the CharMM27 force field for protein and the TIP3P water model for the water molecule[59]. The electrostatic interaction was treated using PME and SHAKE constraints were applied with a 4fs integration time step[60]. The temperature was kept at 300K using a Langevin bath method and the pressure was maintained at 1 ATM using the Berendsen method. The trajectories were analyzed using the MMTSB toolset[61]. For the reaction coordinate (RC) of scatter density distribution, we used the salt-bridge distance between K745 and E762 as one RC. Because the flexibility of Cys797 is low as shown in Figure S4, we measured the two distances from C797 to D855 and F856, where the difference constitutes another RC.

## 4. Function-site interaction fingerprint encoding

Function-site interaction fingerprints (Fs-IFP) are an efficient means to determine functional site binding characteristics and to compare binding sites on a proteome scale as detailed in previous applications[35, 62-63]. In brief, the Fs-IFP method includes three steps. Step 1 prepares the dataset. In this paper, we collected all EGFR kinase-ligand structures as our dataset. Step 2 aligns all function sites. Here we used the SMAP software[52-54] with default parameters. Thus, we obtained all aligned residues within the binding pocket. Step 3 encodes the binding site–ligand interaction. Based on the aligned residues, we confirmed that every binding site could be described using 39 amino acids that comprised the binding pocket. Then the interactions between every amino acid and the corresponding ligand were encoded using a 7-bit array that represents 7 types of interactions: ((1)van der Waals; (2) aromatic face to face; (3) aromatic edge to face; (4) hydrogen bond (protein as hydrogen bond donor); (5) hydrogen bond (protein as hydrogen bond acceptor); (6) electrostatic interaction (protein positively charged); and (7) electrostatic interaction[39]. Finally, the kinase–ligand interactions of each complex structure were encoded using a length of 273 bits (7 bits × 39 residues). The encoding was done using IChem software[64]. Further, the similarity of the pairwise Fs-IFPs was calculated using the Tanimoto coefficient (TC). Based on all-against-all pairwise TC similarity, the hierarchical cluster analysis was carried out using R with the single linkage method[65].

## Supporting information

**Figure S1-9** illustrate the C- and R-spine analysis, the volume and shape of the binding pockets, the analysis of different MD trajectories, PCA, the Class-D1 binding pocket, the different between the Class-D1 binding site and all the other six classes of EGFR binding sites, and the screened top complex. **Table S1-6** describes the EGFR kinase structure dataset, the volume of every sub-pocket conformation, the similarity list, the top similar complex, the index and the pdb id of every kinase structure, and the classes of the binding mode, respectively. The Class-D1 EGFR kinase structure (**PDB**).

# Acknowledgement

We thank Prof. Ajay N. Jain for providing the Surflex Platform version 4.1 license, Prof. Brian Y. Chen for providing VASP software and its application and Prof. Didier Rognan for providing the IChem software. We thank Cameron Mura for helpful discussions. This work was partly supported by the University of Virginia (PEB), Grant Number R01LM011986 from the National Library of Medicine of the National Institute of Health (NIH) (LX), and Grant Number R01GM122845 from the National Institute of General Medical Sciences of NIH (LX).

# Table of Contents Graphic

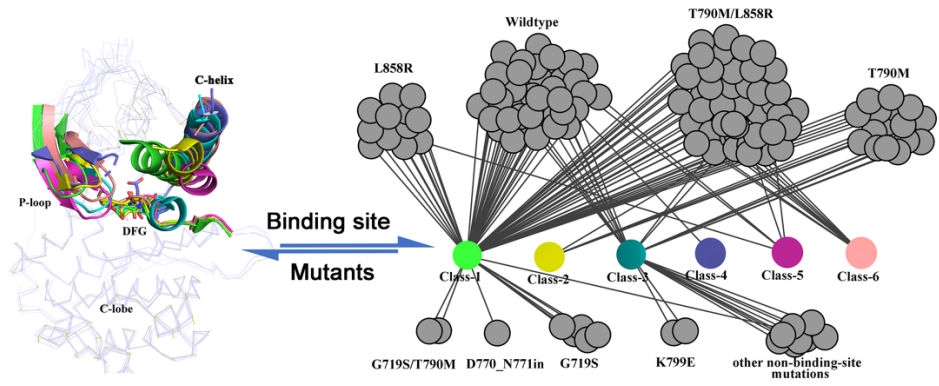